\documentclass[9pt,twocolumn,twoside]{opticajnl}
\journal{ol} % Choose journal (ao,jocn,josaa,josab,ol,optica,pr)

\setboolean{shortarticle}{true}
\usepackage{siunitx}
\title{Needle-beam generation from SPDC photons using inverse-designed flat optics}

\author[1,2]{Sunny}
\author[2,3]{Jyotshna Bora}
\author[2,3]{Ayush Tiwari}
\author[2,3]{C. S. Yadav}
\author[4,5]{Rajesh Menon}
\author[2]{Amit D. Lad}
\author[1,*]{Vishwa Pal}

\affil[1]{Department of Physics, Indian Institute of Technology Ropar, Rupnagar-140001, Punjab, India}
\affil[2]{Center for Quantum Science and Technology, Indian Institute of Technology Mandi, Kamand, Mandi 175075, Himachal Pradesh, India}
\affil[3]{School of Physical Sciences, Indian Institute of Technology Mandi, Kamand, Mandi 175075, Himachal Pradesh, India}
\affil[4]{Department of Electrical and Computer Engineering, University of Utah, Salt Lake City, Utah 84112, USA}
\affil[5]{Oblate Optics, Inc., San Diego, California 92130, USA}

\affil[*]{Corresponding author: vishwa.pal@iitrpr.ac.in}

\begin{abstract}
We experimentally demonstrate the generation of needle-like beams from spontaneous parametric down-conversion (SPDC) light using a compact inverse-designed multilevel diffractive lens (MDL). The nonclassical character of the SPDC source is independently verified through a three-detector heralded Hanbury Brown--Twiss measurement performed before the MDL yielding $g_h^{(2)}(0)\approx 0.14$. Spatial measurements using an intensified charge-coupled device reveal that the MDL maintains a tightly confined focal distribution over an axial distance of approximately $11\thinspace$mm under collinear SPDC illumination, substantially reducing diffraction-induced broadening compared with a conventional plano-convex lens of the same nominal focal length. Under non-collinear SPDC illumination, two distinct regions of extended axial confinement are observed, demonstrating the dependence of the engineered focal distribution on the incident spatial field. These results establish a compact flat-optical approach for engineering extended-focus SPDC photon fields, with potential applications in structured quantum photonics.
\end{abstract}

\setboolean{displaycopyright}{true}

\begin{document}

\maketitle

\section{Introduction}
The ability to control the spatial distribution and propagation of light is fundamental to modern optical technologies, with applications ranging from high-resolution microscopy and optical coherence tomography to laser material processing \cite{Fahrbach2010,Sampson2014,Bhuyan2010}. Optical fields that maintain tight transverse confinement over an extended axial distance are particularly attractive because they can increase the usable depth of focus (DOF) while preserving spatial localization \cite{Zalevsky:10,Zhao:22}. However, conventional refractive lenses generally exhibit a trade-off between focal-spot size and DOF, as strong diffraction-limited focusing is accompanied by rapid beam divergence beyond the focal region \cite{Saleh:07,Grunwald2020}. To overcome this limitation, structured optical fields such as Bessel, Airy, and needle-like beams have been developed to achieve tailored propagation characteristics, including extended axial confinement and reduced diffraction-induced spreading \cite{Durnin:87,Siviloglou:07,Grunwald2020,Chen:24}.

Needle-like beams have been generated using axicons, spatial light modulators, and specially designed diffractive optical elements. Among these approaches, inverse-designed multilevel diffractive lenses (MDLs) offer a compact, lightweight, and planar alternative to conventional optical systems. By engineering the spatially varying phase response, MDLs can provide strong focusing together with an extended DOF \cite{Meem2020,Banerji2020}. Recent studies have demonstrated optical vortex needle beams using flat diffractive lenses \cite{Anita:24} and extended-focus femtosecond laser pulses using MDLs \cite{Jena2025}. Furthermore, optical needle formation with a multilevel diffractive lens has recently been demonstrated for optical coherence tomography \cite{Huang2026}. These developments establish the effectiveness of inverse-designed flat optics for controlling the spatial propagation of classical optical fields.

Extending structured-light concepts to the quantum regime provides opportunities for engineering the spatial degrees of freedom of individual photons. Previous studies have demonstrated spatially entangled airy photons \cite{Lib2020}, self-healing of heralded single-photon airy beams \cite{Li2021}, and high-dimensional structured quantum states \cite{Nape2023,Forbes2025}. Such investigations highlight the potential of spatial-mode engineering in quantum imaging, sensing, and information processing. In this context, combining independently characterized nonclassical photon sources with compact extended-depth-of-focus optical elements provides a route toward structured single-photon photonics.

Here, we experimentally investigate needle-beam formation from spontaneous parametric down-conversion (SPDC) photons using inverse-designed MDLs. The nonclassical character of the SPDC source is independently verified through a three-detector heralded Hanbury Brown--Twiss measurement performed before the MDL, yielding $g_h^{(2)}(0)\approx 0.14$. Using an intensified charge-coupled device (ICCD), we characterize the accumulated spatial distributions of the SPDC field under collinear and non-collinear illumination. For collinear SPDC, an MDL with a nominal focal length of $50\thinspace$mm produces a tightly confined focal distribution extending over approximately $11\thinspace$mm, substantially reducing diffraction-induced broadening compared with a conventional plano-convex lens of the same focal length. Under non-collinear illumination, two distinct regions of enhanced axial confinement are observed. These results demonstrate a compact flat-optical approach for engineering extended-focus spatial distributions of SPDC-generated photons and provide a foundation for structured single-photon beam engineering.
\section{Experimental Setup}\label{sec:experimental}
\begin{figure}[ht!]
\centering
\includegraphics[width=\linewidth]{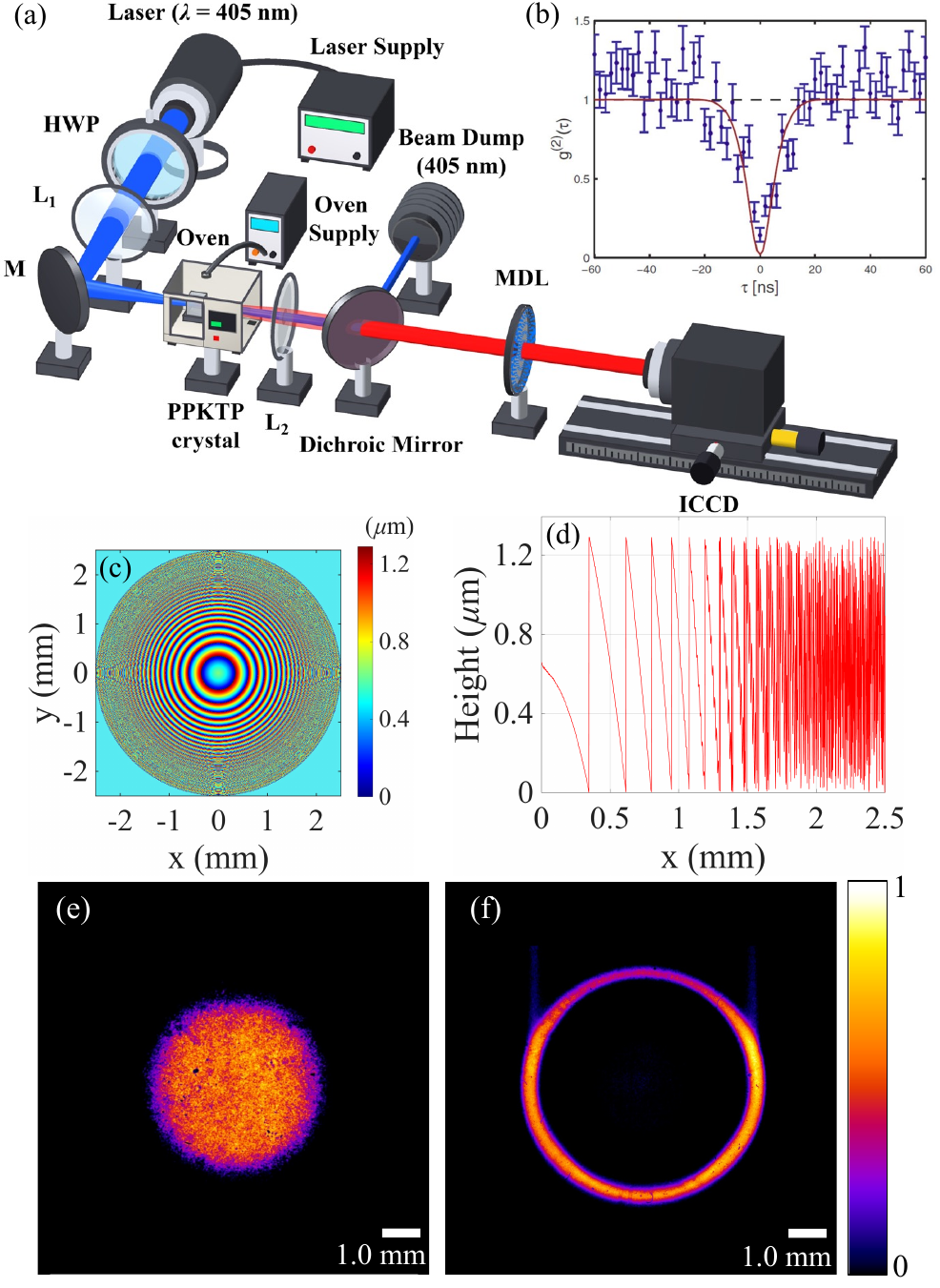}
\caption{Experimental setup for generating needle-like beams from SPDC photons using an inverse-designed MDL. (a) Schematic of the experimental arrangement. HWP: half-wave plate; L$_{1}$ and L$_{2}$: plano-convex lenses; PPKTP: periodically poled potassium titanyl phosphate crystal; M: mirror (M); MDL: multilevel diffractive lens. (b) Heralded second-order correlation function, $g_h^{(2)}(\tau)$, measured before the MDL. Black squares denote experimental data, and the red solid curve represents the fit, yielding $g_h^{(2)}(0)\approx 0.14$. (c) Designed height profile of the MDL and (d) it's cross-section. Normalized spatial distributions of the incident SPDC field for (e) collinear and (f) non-collinear configurations.}
\label{ExpSch} 
\end{figure}

The experimental schematic for generating needle-like beams from SPDC photons is shown in Fig.\,\ref{ExpSch}(a). A pump laser operating at a wavelength of $\lambda=405\thinspace$nm passes through a half-wave plate (HWP) for polarization control and is subsequently focused into a periodically poled potassium titanyl phosphate (PPKTP) crystal using a plano-convex lens, L$_1$, with a focal length of $f_1=200\thinspace$mm. The crystal is mounted in a temperature-controlled oven to achieve the desired phase-matching conditions. Through the SPDC process, the pump generates correlated signal and idler photons with wavelengths near $810\thinspace$nm under nominally degenerate conditions. Both collinear and non-collinear SPDC configurations are investigated.

The down-converted photons are collimated using plano-convex lens L$_2$ ( $f_2=50\thinspace$mm), while a dichroic mirror separates the residual $405\thinspace$nm pump from the SPDC field. The resulting SPDC photons are directed onto an inverse-designed MDL, which controls their spatial distribution and axial focusing characteristics. The transverse intensity distribution is recorded at different propagation distances using an ICCD camera. The ICCD is operated without herald-triggered gating, with an exposure time of $0.3\thinspace$s. Consequently, the recorded images represent accumulated spatial distributions of the SPDC field rather than directly measured herald-conditioned photon distributions.

To independently characterize the nonclassical nature of the photon source, a three-detector heralded Hanbury Brown--Twiss (HBT) measurement is performed before the MDL. One photon of the SPDC pair serves as the herald, while the correlated photon is directed into a two-detector HBT arrangement. The measured heralded second-order correlation function, $g_h^{(2)}(\tau)$, is shown in Fig.\,\ref{ExpSch}(b). A pronounced suppression at zero time delay yields $g_h^{(2)}(0)\approx0.14$, indicating a substantial suppression of multiphoton contributions in the heralded photon state.

The MDL is designed using an inverse method \cite{Meem:20,Molesky:18}. Its complex radial structure comprises thousands of concentric rings with spatially varying widths and heights, which impart position-dependent phase retardation to the incident wavefront. The MDL is fabricated on a polycaronate substrate via UV casting based replication (Edmund Optics Stock $\#$90-147). The primary MDL has an aperture diameter of $D=10\thinspace$mm, a nominal focal length of $f=50\thinspace$mm, and a design wavelength of $808\thinspace$nm. The MDL is designed for plane-wave illumination, whereas the experimental SPDC illumination has a finite transverse spatial extent. The height profile and corresponding radial cross-sectional profile of the MDL are shown in Figs.\,\ref{ExpSch}(c) and \ref{ExpSch}(d), respectively. A second MDL with a focal length of $f=200\thinspace$mm is also investigated, with the corresponding results are presented in Supplement\,1.

The SPDC photons are generated in collinear and non-collinear configurations. Under collinear phase matching, the signal and idler photons propagate approximately along the pump-beam direction, producing a nearly circular spatial distribution. In the non-collinear configuration, the down-converted photons propagate at finite angles relative to the pump axis, producing an annular distribution under approximately azimuthally symmetric conditions. The corresponding spatial distributions are shown in Figs.\,\ref{ExpSch}(e) and \ref{ExpSch}(f), respectively.
%Figures\,\ref{ExpSch}(e) and \ref{ExpSch}(f) show the normalized spatial distributions of the incident SPDC field for collinear and non-collinear configurations, respectively. 
The collinear configuration produces an approximately circular spot with a diameter of $5\thinspace$mm, whereas the non-collinear configuration yields an annular distribution with a ring diameter of approximately $6.4\thinspace$mm.

In the experiments, the incident pump power at $405\thinspace\mathrm{nm}$ is maintained within the range of $P_p=7$–$10\thinspace\mathrm{mW}$. The crystal temperature is set to $T_c=50,^{\circ}\mathrm{C}$ for collinear SPDC and $T_c=55,^{\circ}\mathrm{C}$ for noncollinear SPDC configurations.

%For the collinear SPDC, the operating parameters were an incident $405\thinspace$nm pump power of $P_p=7\thinspace$mW, a crystal temperature of $T_c=50\,^{\circ}\mathrm{C}$, and an ICCD gain setting of 2000. 

%For the noncollinear SPDC, the experimental parameters were an incident $405\thinspace$nm pump power of $P_p=10\thinspace$mW, crystal temperature $T_c=55\,^{\circ}\mathrm{C}$, and ICCD gain of 2800 for the MDL, and $P_p=7.5\thinspace$mW, $T_c=45\,^{\circ}\mathrm{C}$, and ICCD gain of 2200 for the PCXL.
%
\section{Results and Discussion}

\subsection{Needle-beam formation under collinear SPDC illumination}
\begin{figure}[htbp]
\centering
\includegraphics[width=\linewidth]{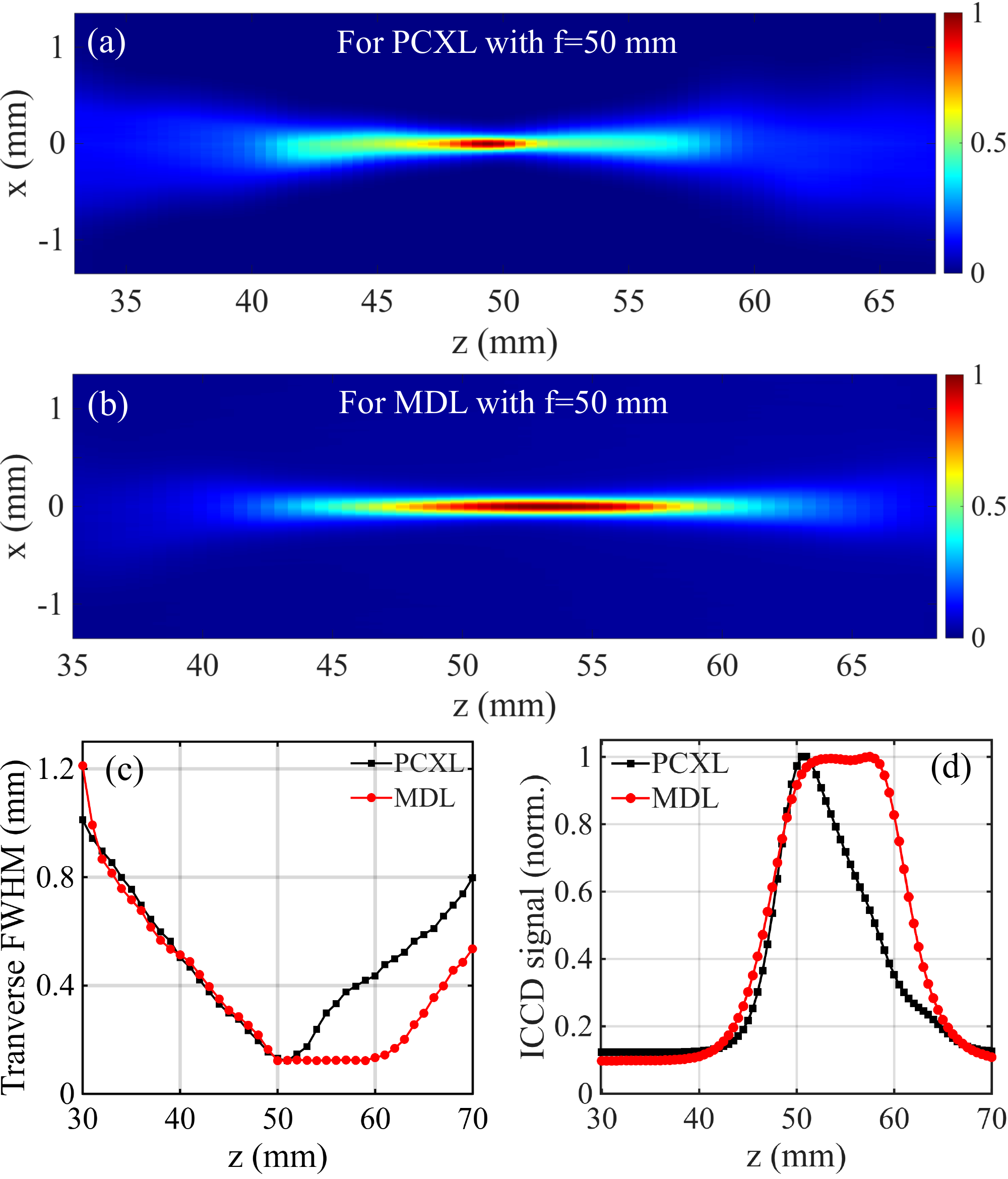}
\caption{Evolution of the spatial distribution of collinear SPDC photons in the $x$--$z$ plane for an incident beam diameter of $5\thinspace$mm using (a) a conventional PCXL and (b) an MDL, both having a nominal focal length of $f=50\thinspace$mm. (c) Variation of the transverse focal-spot size and (d) on-axis ICCD signal as functions of axial propagation distance. The black curve with filled squares and the red curve with filled circles represent the results obtained using the conventional PCXL and MDL, respectively.}
\label{fig3} 
\end{figure}

First, we investigate the focusing characteristics of collinear SPDC photons using the MDL and compare the results with those obtained using a conventional plano-convex lens (PCXL). An input SPDC beam with a diameter of $5\thinspace$mm, as shown in Fig.\,\ref{ExpSch}(e), is directed onto either the MDL or PCXL. Both MDL and PCXL have a nominal focal length of $f=50\thinspace$mm. The corresponding experimental results are presented in Fig.\,\ref{fig3}. 

Figures~\ref{fig3}(a) and \ref{fig3}(b) show the measured spatial evolution of the SPDC field in the $x$--$z$ plane for the PCXL and MDL, respectively. The axial propagation distance $z$ is measured from the plane of the respective optical element. For the PCXL, the incident field converges toward a minimum transverse width near $z=50\thinspace$mm and subsequently diverges due to diffraction. In contrast, the MDL maintains a narrow transverse distribution over a substantially extended axial region, resulting in a needle-like focal structure.

%Figures\,\ref{fig3}(a) and \ref{fig3}(b) show the evolution of the photon distribution as a function of propagation distance for the conventional plano-convex lens and the inverse-designed MDL, respectively. Both optical elements have the same focal length of $5\thinspace$cm. The propagation distance is measured from the plane of the respective optical element.

%First, we have investigated the focusing characteristics of collinear single photons using the inverse-designed MDL and compared them with those obtained using a conventional plano-convex lens. For the experiment, we considered an input single-photon beam with a diameter of $5\thinspace$mm, as shown in Fig.\,\ref{fig2}(a), and directed it onto either MDL or a plano-convex lens. The results are shown in Fig.\,\ref{fig3}. Figures\,\ref{fig3}(a) and \ref{fig3}(b) shows the evolution of photon counts as a function of propagation distance for the cases of an MDL and a conventional plano-convex lens of the same focal length of $5\thinspace$cm, respectively. Note, the propagation distances are measured with respect to MDL/plano-convex lens plane.
To quantify the focusing characteristics, we determine the transverse full width at half maximum (FWHM) and the on-axis ICCD signal as functions of propagation distance, as shown in Figs.\,\ref{fig3}(c) and \ref{fig3}(d), respectively. The transverse width is obtained from the FWHM of the recorded spatial distribution at each axial position. While both optical elements produce a minimum focal width of the order of $100\thinspace\mu$m, the PCXL (black curve with filled squares) exhibits rapid transverse broadening away from its focal plane. In contrast, the MDL (red curve with filled circles) maintains a nearly constant transverse width over an axial range of approximately $11\thinspace$mm, demonstrating its extended-depth-of-focus response.

The corresponding axial ICCD signals are presented in Fig.\,\ref{fig3}(d). The PCXL exhibits a localized maximum near the nominal focal plane, followed by a rapid decrease in the recorded on-axis signal. The MDL, however, produces a broader axial response, with the on-axis signal remaining relatively high over the extended focal region. The simultaneous persistence of a narrow transverse distribution and an extended axial response confirms the formation of a needle-like spatial distribution in the SPDC field.

These observations establish that the inverse-designed MDL can substantially modify the axial focusing characteristics of SPDC photons while maintaining tight transverse confinement. The resulting extended focal region provides a compact flat-optical approach for spatial beam shaping in quantum regime, overcoming the localized focal response of conventional refractive lenses.

\subsection{Needle-beam formation under non-collinear SPDC illumination}
\begin{figure}[ht!]
\centering
\includegraphics[width=\linewidth]{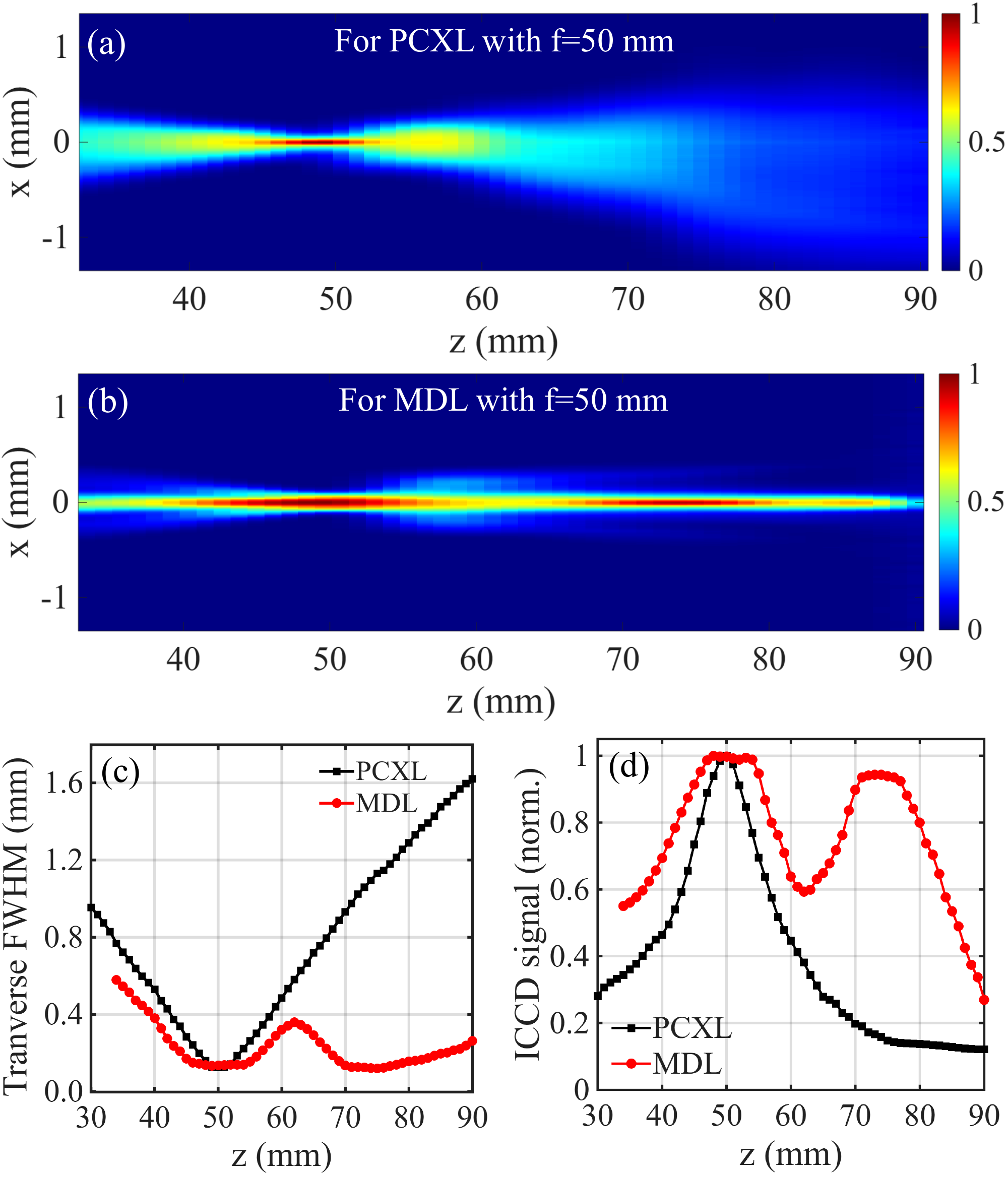}
\caption{Evolution of the spatial distribution of non-collinear SPDC photons in the $x$--$z$ plane using (a) a conventional PCXL and (b) an MDL, both having a nominal focal length of $f=50\thinspace$mm. The incident SPDC field has an annular spatial distribution with a ring diameter of approximately $6.4\thinspace$mm. (c) Variation of the transverse focal-spot size and (d) on-axis ICCD signal as functions of propagation distance. The black curve with filled squares and the red curve with filled circles represent the results obtained using the conventional PCXL and MDL, respectively.}
\label{fig5} 
\end{figure}

We further investigate the focusing characteristics of non-collinear SPDC photons using the MDL. In this configuration, the incident SPDC field exhibits an annular spatial distribution, as shown in Fig.\,\ref{ExpSch}(e), with a ring diameter of approximately $6.4\thinspace$mm. The focusing characteristics are compared qualitatively with those of a conventional PCXL having the same nominal focal length of $f=50\thinspace$mm. The experimental results are presented in Fig.\,\ref{fig5}.

Figure\,\ref{fig5}(a) shows the measured spatial evolution of the non-collinear SPDC field using the conventional PCXL. The field converges toward a localized focal region near $z=50$~mm and subsequently undergoes transverse broadening. In contrast, the MDL produces a more complex axial focusing response, as shown in Fig.\,\ref{fig5}(b), with narrow transverse distributions extending over distinct axial regions.

The corresponding transverse FWHM and on-axis ICCD signals are presented in Figs.\,\ref{fig5}(c) and \ref{fig5}(d), respectively. While the conventional lens exhibits a single dominant focal region, the MDL produces two regions of enhanced axial confinement. The on-axis ICCD signal exhibits two local maxima, approximately around $z=50\thinspace$mm and $z=75\thinspace$mm, accompanied by reduced transverse widths over the corresponding axial intervals. These observations reveal a nontrivial focusing response of the MDL under annular SPDC illumination.

The appearance of two axial confinement regions may be influenced by the annular spatial distribution of the incident SPDC field and the wavelength-dependent phase response of the diffractive lens. However, the present measurements do not uniquely establish their physical origin.  Furthermore, the MDL and PCXL measurements were performed at different incident pump powers and crystal temperatures. For the MDL, the pump power was $P_p=10\thinspace$mW and the crystal temperature was $T_c=55\,^{\circ}\mathrm{C}$, whereas for the PCXL the corresponding values were $P_p=7.5\thinspace$mW and $T_c=45\,^{\circ}\mathrm{C}$. Since the crystal temperature influences the SPDC phase-matching conditions, these differences may affect the spectral and angular distributions of the incident SPDC field. Consequently, the comparison is restricted to the observed spatial focusing characteristics, and differences in absolute ICCD signal levels are not interpreted as changes in optical efficiency.

These results illustrate the sensitivity of the extended-focus response of inverse-designed flat optics to the spatial characteristics of the incident SPDC field. Together with the collinear measurements, they demonstrate the ability of the MDL to produce different needle-like axial distributions under distinct SPDC illumination conditions.

Additional measurements using a 10~mm-diameter collinear SPDC beam, together with focusing results obtained using a second MDL with a nominal focal length of $200\thinspace$mm, are provided in Supplement\,1.

The spatial transformation produced by the MDL is governed by passive linear optical propagation and is therefore applicable to individual photon spatial modes. Accordingly, the observed extended-focus distributions provide information relevant to the engineering of needle-like single-photon modes. The independent pre-MDL heralded correlation measurement establishes the nonclassical character of the SPDC source, whereas the ICCD measurements characterize the accumulated spatial response. Although herald-conditioned spatial distributions were not directly recorded after the MDL, the combined observations provide an experimental foundation for developing compact flat-optical elements for structured single-photon photonics.
%We have also performed experiments with another MDL having different design parameters of focal distance, focused spot size and EDOF. We obtained focused fs-pulses with an average spot size of $\sim60\thinspace\mu$m over an extended range of propagation distances  $z=27\thinspace\mathrm{cm}-37\thinspace\mathrm{cm}$ (see supplement 1). Further, the on-axis intensity within EDOF is found to vary within $40-100\%$. The maximum $\eta$ is found to be $7\%$ (see supplement 1).

%The lower focusing efficiencies arise from fabrication errors in MDLs, misalignments, and imperfections in the fs-laser beam profile, as the MDLs were designed using an inverse method assuming uniform plane-wave illumination. Additionally, discrepancies between the 2nd design (1064 nm) and experimental wavelength (800 nm, 60 nm bandwidth) further reduce efficiency. This work demonstrates the feasibility of using flat MDLs for generating focused fs-pulses, in plasma channel formation and micromachining, with future efforts aimed at significantly improving focusing efficiency.
%\vspace{-6pt}
\section{Conclusion}
In conclusion, we have experimentally demonstrated the generation of needle-like extended-focus distributions of SPDC photons using compact, inverse-designed MDLs. Under collinear SPDC illumination, the MDL maintains a tightly confined transverse distribution over an axial range of approximately $11\thinspace$mm, substantially reducing diffraction-induced broadening compared with a conventional PCXL of the same nominal focal length. Under non-collinear SPDC illumination, two distinct regions of enhanced axial confinement are observed, highlighting the dependence of the engineered focal distribution on the incident spatial field. The nonclassical character of the SPDC source is independently verified through a pre-MDL heralded second-order correlation measurement, yielding $g_h^{(2)}(0)\approx0.14$. While the spatial distributions are recorded without herald-conditioned detection, the observed focusing characteristics demonstrate the capability of inverse-designed flat optics to engineer extended-focus SPDC fields. These results provide a compact and versatile approach toward structured single-photon photonics, with potential applications in quantum imaging, quantum sensing, quantum communication, and light--matter interactions.

\begin{backmatter}
\bmsection{Funding} 
National Quantum Mission (NQM) of the Department of Science and technology, Government of India; Department of Biotechnology (DBT), GOI, under Project Grant No.~BT/PR65032/QTBT/176/26/2025; and the IIT Mandi Seed Grant (No.~IITM/SG/ADL/173).

\bmsection{Acknowledgments} Sunny thanks IIT Ropar for the financial support. JB and AT acknowledge financial support from the IIT Mandi HTRA Fellowship.

\bmsection{Disclosures} RM: Oblate Optics (I,E,P). 

\bmsection{Data availability} Data underlying the results presented in this paper are not publicly available at this time but may be obtained from the authors upon reasonable request.

\bmsection{Supplemental document}
See Supplement 1 for supporting content. 

\end{backmatter}

% Bibliography
%\bibliography{sample}

%\bibliographyfullrefs{sample}

\end{document}